\RequirePackage{snapshot}
\documentclass[sigconf,table,letter]{acmart}

\usepackage[utf8]{inputenc}
\usepackage{graphicx}
\usepackage{subfig}
\usepackage{xcolor}  
\usepackage{amsmath,amssymb}
\usepackage{ifthen}
\usepackage{xspace}
\usepackage{flushend}
\usepackage{multicol}
\usepackage{enumitem}
\setlist{leftmargin=5.5mm}
\usepackage[ruled,vlined, linesnumbered]{algorithm2e}
\usepackage{algpseudocode}
\usepackage{color}
\usepackage{listings}
\usepackage{pifont}
\usepackage{multirow}
\usepackage{url}
\renewcommand*{\UrlFont}{\ttfamily\rm}

\newcommand{\peas}{PEAS\xspace}
\newcommand{\xsearch}{\textsc{X-S\small{earch}}\xspace}
\newcommand{\tor}{Tor\xspace}
\newcommand{\direct}{Direct\xspace}
\newcommand{\tmn}{TrackMeNot\xspace}
\newcommand{\goopir}{GooPIR\xspace}
\newcommand{\queryscrambler}{QueryScrambler\xspace}

\usepackage{listings}

\newcommand{\SYS}{\xsearch}

\newboolean{showcomments}
\setboolean{showcomments}{true}
\ifthenelse{\boolean{showcomments}}
{ \newcommand{\mynote}[3]{
   \fbox{\bfseries\sffamily\scriptsize#1}
   {\small$\blacktriangleright$\textsf{\emph{\color{#3}{#2}}}$\blacktriangleleft$}}}
{ \newcommand{\mynote}[3]{}}

\definecolor{darkgreen}{rgb}{0.3,0.5,0.3}
\definecolor{darkblue}{rgb}{0.3,0.3,0.5}
\definecolor{darkred}{rgb}{0.5,0.3,0.3}

\copyrightyear{2017}
\acmYear{2017}
\setcopyright{othergov}
\acmConference{Middleware '17}{December 11--15, 2017}{Las Vegas, NV,
USA}\acmPrice{15.00}\acmDOI{10.1145/3135974.3135987}
\acmISBN{978-1-4503-4720-4/17/12}

\begin{document}

\title{\textsc{X-Search}\xspace: Revisiting Private Web Search using Intel SGX}
\date{}

\author{Sonia Ben Mokhtar}
\affiliation{  \institution{LIRIS, CNRS, University of Lyon}
  \city{Lyon}
  \country{France}
}
\email{sonia.benmokhtar@liris.cnrs.fr}

\author{Antoine Boutet}
\affiliation{  \institution{Univ. of Lyon, Inria, INSA Lyon, CITI}
  \city{Lyon}
  \country{France}
}
\email{antoine.boutet@insa-lyon.fr}

\author{Pascal Felber}
\affiliation{  \institution{University of Neuchâtel}
  \city{Neuchâtel}
  \country{Switzerland}
}
\email{pascal.felber@unine.ch}

\author{Marcelo Pasin}
\affiliation{  \institution{University of Neuchâtel}
  \city{Neuchâtel}
  \country{Switzerland}
}
\email{marcelo.pasin@unine.ch}

\author{Rafael Pires}
\affiliation{  \institution{University of Neuchâtel}
  \city{Neuchâtel}
  \country{Switzerland}
}
\email{rafael.pires@unine.ch}

\author{Valerio Schiavoni}
\affiliation{  \institution{University of Neuchâtel}
  \city{Neuchâtel}
  \country{Switzerland}
}
\email{valerio.schiavoni@unine.ch}

\renewcommand{\shortauthors}{S. Ben Mokhtar, A. Boutet, P. Felber, M. Pasin, R. Pires, and V. Schiavoni}

\begin{abstract}
The exploitation of user search queries by search engines is at the heart of their economic model.
As consequence, offering private Web search functionalities is essential to the users who care about their privacy. 
Nowadays, there exists no satisfactory approach to enable users to access search engines in a privacy-preserving way. 
Existing solutions are either too costly due to the heavy use of cryptographic mechanisms (e.g., private information retrieval protocols), subject to attacks (e.g., Tor, TrackMeNot, GooPIR) or rely on weak adversarial models (e.g., PEAS). 
This paper introduces \SYS, a novel private Web search mechanism building on the disruptive Software Guard Extensions (SGX) proposed by Intel. 
We compare \SYS to its closest competitors, Tor and PEAS, using a dataset of real web search queries. 
Our evaluation shows that: (1) \SYS offers stronger privacy guarantees than its competitors as it operates under a stronger adversarial model; (2) it better resists state-of-the-art re-identification attacks; and (3) from the performance perspective, \SYS outperforms its competitors both in terms of latency and throughput by orders of magnitude.
\end{abstract}


\maketitle

\section{Introduction}
\label{sec:introduction}
\vskip 5mm
Web search is with no doubt the most widely used online service, with more than 3.5 billion queries sent on a daily basis to Google alone.
These queries are generally stored by search engines to analyze user behavior and to personalize responses according to profiles inferred from the past queries of the users~\cite{langville2011google,Hannak:2013:MPW:2488388.2488435}.
They are at heart of the economic model of online services, which heavily relies on (personalized) advertising~\cite{yang2010analyzing}.
However, as pointed out by numerous studies, the collection and exploitation of search queries opens a number of privacy threats as they can disclose sensitive information about individuals (e.g., their age, sex, religious or political preferences, sexual orientation)~\cite{castelluccia2010private}.

To deal with this issue, a number of solutions enabling the users to query search engines in a privacy preserving manner have been proposed in the literature.
These solutions can be classified in three categories according to the guarantees they offer to the users.

The first category of solutions are those enforcing \emph{unlinkability} between a user and her search query.
The most popular approaches in this category are anonymous communication protocols (e.g., Tor~\cite{dingledine2004tor}, Dissent~\cite{corrigan2010dissent,wolinsky2012dissent}, RAC~\cite{mokhtar2013rac}).
These solutions are however limited for two main reasons:
first, they typically suffer from poor performance because of the heavy cryptographic mechanisms they rely on;
second, despite ensuring anonymity of the requester, it has been shown in~\cite{peddinti2014web} that the actual content of search queries may be sufficient to link back to the identity of the user.

To overcome this limitation, a second category of solutions aim at enforcing \emph{indistinguishability} between user profiles/queries.
To that end, they obfuscate user preferences/profile in such a way that the search engine cannot distinguish between a user's real interests and fake ones (e.g., Track me not~\cite{howe2009trackmenot}, GooPIR~\cite{domingo2009h}).
These approaches generally operate by sending fake queries (also called dummy queries) on behalf of the user.
It has been shown~\cite{petit2016simattack}, however, that the external resources used for generating fake queries (e.g., RSS feeds, dictionaries) makes it possible for search engines to easily distinguish fake from real traffic.
Combination of unlinkability and indistinguishability has also been proposed in the literature, yet the only existing solution that we are aware of (PEAS~\cite{petit2015peas}) assumes a weak adversarial model of non-colluding proxy servers.

The last category of solutions are those enabling private information retrieval (PIR), e.g., \cite{pang2010privacy,lindell2010private}).
These approaches rely on specialized search engines implementing cryptographic techniques (e.g., homomorphic encryption) that enable to answer a user request without having access to its content.
These techniques are, however, still unpractical due to their limited performance with response times in the order of seconds for very large data stores~\cite{aguilar2016xpir}, which is the case of search engines. 

Based on these considerations, it appears clearly that to fully support privacy-preserving Web search one must address two main challenges.
The first one is to provide a practical and secure unlinkability protocol, i.e., a protocol enabling the protection of the identity of the requester in a more realistic adversarial model, without compromising the interactiveness between the user and the search engine.
The second one is to provide an effective indistinguishability protocol that generates realistic fake queries, i.e., difficult to distinguish from real queries.

This paper contributes \SYS{}, a novel privacy proxy enabling Internet users to access Web search engines in a privacy-preserving manner.
\SYS{} relies on Intel software guard extensions (SGX)~\cite{costan_intel}, a hardware technology that provides a trusted execution environment able to perform secure computations within an ``enclave''.
Instead of submitting her queries directly to the search engine, a user sends them to the \SYS{} proxy to execute them on her behalf. 
The proxy executes attested code in a trusted SGX enclave (see \S\ref{subsec:sgx} for details on the guarantees provided by SGX).
The queries are encrypted while outside the enclave, and only accessible as plain text from within.
The \xsearch proxy then generates an obfuscated query by aggregating $k$ random past queries and the original one using the logical OR operator in such a way that the search engine is not able to distinguish which one is the original query.
As the obfuscated scheme can alter the results returned by the search engine by mixing results for the original query with results for the additional aggregated past queries, the \xsearch proxy filters results to only forward to the user the results related to the initial query.

We evaluate \SYS{} from three perspectives: privacy, accuracy, and performance.
From the privacy perspective, we analytically show that \SYS{} offers stronger privacy guarantees than its competitors as it operates under a stronger adversarial model.
Furthermore, we experimentally demonstrate using a data set of real search queries that \SYS{} is more resilient to state-of-the-art re-identification attacks than \peas (by $30\%$ in average).
From the accuracy perspective, we show that the impact of the obfuscation scheme of \xsearch remains limited. 
For instance, with two fake queries in the obfuscated query, the user retrieves more than $80\%$ of the results returned for the initial query.
From the performance perspective, we show that \SYS{} outperforms its competitors both in terms of latency and throughput. Specifically, the throughput of \SYS{} is one order of magnitude higher than the one of \peas and two orders of magnitude higher than the one of \tor.

The contributions of \SYS{} are as follows. First, we present a novel architecture to allow privacy-preserving Web searches that exploits Intel SGX to operate under stronger adversarial models than existing systems in literature.
Second, we contribute a novel query obfuscation mechanism.
Third, we present the implementation choices of our full prototype. 
Finally, we contribute an extensive evaluation, both analytically and experimentally using real-world datasets.

The remainder of the paper is organized as follows.
We first introduce background concepts and overview related work in Section~\ref{sec:relwork}.
Then we present the considered adversary model in Section~\ref{sec:adversary} before presenting our \xsearch proposed protocol in Section~\ref{sec:xsearch}.
Finally, we describe the considered experimental setup and the evaluation of \xsearch in Section~\ref{sec:experiment} and Section~\ref{sec:evaluation}, respectively.
Section~\ref{sec:conclusion} presents our conclusions.

\section{Background and related work}
\label{sec:relwork}

We start by describing in this section the related work in private Web search (Section~\ref{subsec:related-search}).
Then, we discuss the limitations of existing solutions (Section~\ref{subsec:issues}). 
Finally, we present Intel software guard extensions (SGX) and discuss how this novel technology can be used to improve the state of research in the field of private Web search (Section~\ref{subsec:sgx}).

\subsection{Private Web Search}
\label{subsec:related-search}

Private Web search has been an active research area in the last decade in order to counterbalance the numerous threats open due to the oversharing of users' search queries by search engines. 
This research field is likely to gain even more attention due to the recent legislation change in the United States, which enable ISPs to sell user browsing history without their consent.\footnote{http://www.nbcnews.com/news/us-news/trump-signs-measure-let-isps-sell-your-data-without-consent-n742316} 
In this context, existing solutions to private Web search can be classified in three main categories. 
The first two categories (presented respectively in Sections~\ref{subsubsec:unlink} and \ref{subsubsec:indist}) enable clients to use existing search engines while offering them additional privacy guarantees. 
The third category (see in Section~\ref{subsubsec:other}) includes alternative search engines implementing specific privacy-preserving protocols.

\subsubsection{Enforcing unlinkability}
\label{subsubsec:unlink}

This category of solutions includes a set of protocols enabling users to send their search queries anonymously to a search engine, thus enforcing unlinkability between the user identity (e.g., IP address) and her query.

The most popular protocol among these solutions is \tor~\cite{dingledine2004tor}, an implementation of the Onion Routing protocol~\cite{goldschlag1999onion}. 
Similarly to Onion Routing, \tor sends each query through multiple nodes using a cryptographic protocol. 
In this protocol, queries are encrypted using multiple keys of randomly selected nodes (creating an ``onion'' with multiple layers) and routed through these nodes. 
Then, each node deciphers the received cipher text (hence removing the outer-most layer of the onion) and forwards it to the next node until the onion reaches the exit node. 
The exit node retrieves the query and sends it to the search engine on behalf of the user. 
This protocol assumes the participating relays to faithfully forward the onions, which might not be true as some may behave selfishly (e.g., by dropping onions) or even maliciously (e.g., by injecting fake traffic to slow down the system).

RAC~\cite{mokhtar2013rac} overcome these limitations, by enabling anonymous communication in presence of malicious and selfish nodes. 
In this protocol, nodes are organized on several virtual rings such that, for a given ring, a node has a predecessor node and a successor node. 
A node might be part of several rings and thus have multiple predecessors and successors. 
To ensure that no message is dropped by a freerider, nodes have to broadcast all messages they relay. 
Broadcast messages have to circulate through all nodes in the ring such that if a node does not receive a message from one of its predecessors, it considers this predecessor as a freerider. 
The modifications made by RAC suffer from performance limitations, achieving a throughput that is orders of magnitude lower than \tor.

Another robust solution to anonymous communication is the Dissent protocol~\cite{corrigan2010dissent,wolinsky2012dissent}. 
This protocol enforces accountability in presence of malicious and selfish participants. 
However, its performance is even worse than the one of RAC as it is a combination of two heavy cryptographic protocols: the dining cryptographers protocol (DC-NET)~\cite{chaum1988dining} and a data mining protocol used to permute a set of fixed-length messages with cryptographically strong anonymity~\cite{brickell2006efficient}.

In addition to the performance issue, protocols enforcing unlinkability have also been shown not to resist re-identification attacks~\cite{petit2016simattack}. Indeed, the issue comes from the fact that search queries themselves disclose enough information for breaking the unlinkability property. 

\subsubsection{Enforcing indistinguishability}
\label{subsubsec:indist}

To protect users against re-identification attacks, solutions enforcing indistinguishability have been proposed. 
The aim of these solutions is to avoid search engines distinguishing between a user's real interests and fake ones, hence protecting her privacy. 
This is generally achieved either by generating fake queries (e.g., \tmn~\cite{howe2009trackmenot}, \goopir~\cite{domingo2009h}) or by altering the user's query (e.g., \queryscrambler~\cite{arampatzis2013query}). 

\tmn is a Firefox plugin that periodically generates fake queries and send them to the search engine on behalf of the user and independently of her real queries. 
Fake queries in \tmn are generated using RSS feeds.

\goopir introduces $k$ fake queries inside the user's real query. 
All these queries (i.e., the real one and the $k$ fake ones) are separated by the logical \texttt{or} operator and sent to the search engine. 
Fake queries in \goopir are generated by using randomly selected keywords from a dictionary.

\queryscrambler protects users by replacing their queries by semantically related queries. More precisely, for each user query, it generates a set of related queries by generalizing the concepts used in the initial query. Then, by merging and filtering all the results obtained with these related queries, it retrieves the most plausible results for the initial query.

\peas improves over existing solutions by combing an unlinkability protocol with an indistinguishability protocol. 
The former is based on two non-colluding proxy servers. The first one handles user identities without having access to their requests, while the second generates fake queries, and send them to the search engine on behalf of the user. To generate fake queries, \peas uses a co-occurrence matrix built from past user queries. 

\begin{figure}[t!]
\centering
\includegraphics[scale=0.73,trim=0 0 0 0]{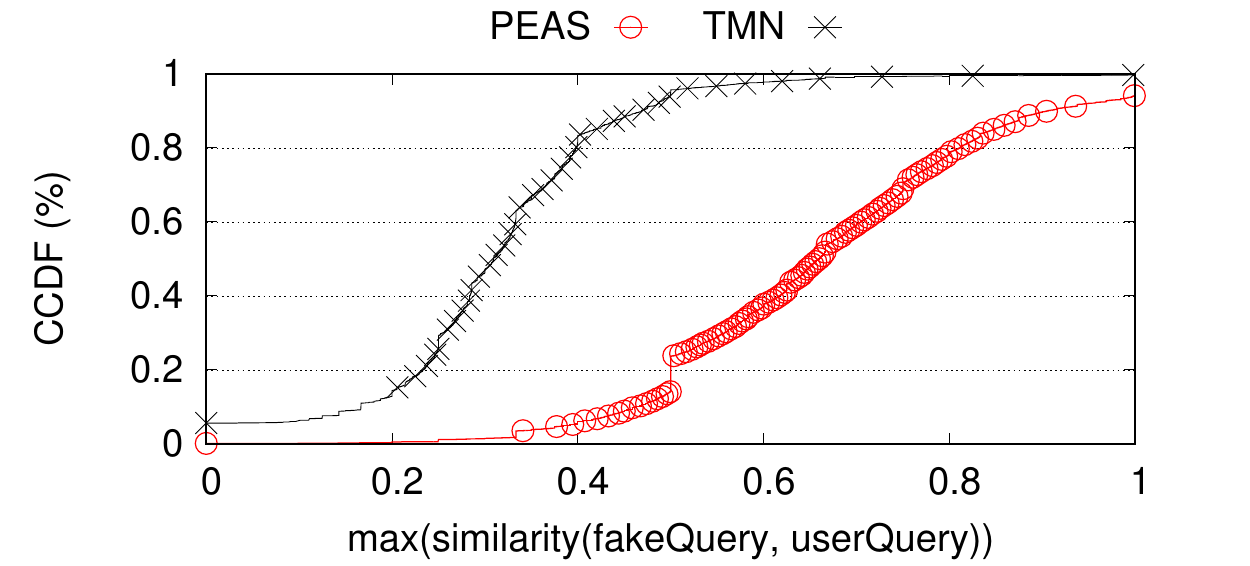}
\caption{Generation of efficient fake queries: almost all fake queries built by \tmn and \peas are original, i.e. never appear in the AOL. CCDF=Complementary Cumulative Distribution Function.}
\label{fig:sim}
\end{figure}

One of the major limitation of these solutions is that it is still easy to discern the fake queriesfrom real ones, as shown by re-identification attacks~\cite{petit2016simattack}. 
We highlight this issue in Figure~\ref{fig:sim}.
The show the CCDF (i.e., Complementary Cumulative Distribution Function) of the maximum similarity between fake queries generated by \peas (i.e., based on the co-occurrence of terms in past queries) and \tmn (i.e., based on RSS feeds) and past queries on the AOL dataset (see Section~\ref{sec:experiment:metrics} to have details of the used dataset and similarity metric). 
This result shows that in both cases most of the fake queries are significantly different from real queries.

\subsubsection{Alternative Search Engines}
\label{subsubsec:other}

This category of solutions build alternative search engines generally based on Private Information Retrieval (PIR) thus enforcing privacy-by-design. 
In these systems, users access information stored on the distant server without revealing to the latter what information they access. 
The only information known by the search engine is that the user has sent a query. 
In general, PIR protocols consist of three algorithms: the constructions of protected queries (keywords are at least encrypted), the execution of the information retrieval (preventing the search engine to access the query and its results), and finally the reconstruction of the result list. Part of these algorithms is performed on the clients, the other part on the distant server.
These generally rely on heavy and unpractical~\cite{aguilar2016xpir} cryptographic protocols, especially when the accessed data stores contain millions of documents, the normal case for today's search engines.

\subsection{Open Challenges in Private Web Search}
\label{subsec:issues}

From the analysis of state of art private Web search solutions, we distinguish two major challenges: one for enforcing unlinkability and one for enforcing indistinguishability. 
The main open challenge for enforcing unlinkability is to design efficient protocols that resist strong adversaries. Indeed, existing protocols are either efficient but assume honest but curious servers (e.g., \tor, \peas) or robust to malicious adversaries but have unpractical performance (e.g., Dissent, RAC).

In term of indistinguishability, the main open challenge is to better resist re-identification attacks by effectively hiding the original query among fake queries. 
This requires the generation of realistic fake queries that are as close as possible to real queries. 

The remaining of this section shows how to leverage Intel Software Guard Extensions and address the above two challenges.

\subsection{Intel Software Guard Extensions}
\label{subsec:sgx}

Cloud software runs in multitenant computing nodes, remotely maintained by third parties.
From the clients' point of view, the environment to remotely run their software can be compromised in several ways.
The third party or the person in charge of managing its hardware may be malicious.
System managers have total access privileges on their hardware to potentially access or tamper with any stored information.
Besides, the remote machine may run compromised operating systems, possibly executed by another (malicious) tenant.
It is therefore hard to trust software running in the clouds.

Homomorphic encryption \cite{Gentry:2009:FHE:1536414.1536440} is an appealing solution for untrusted environments.
A user encrypts data, send it to an untrusted server. 
It is still able to process the ciphertext without having access to its content.
The algorithms proposed so far prove that the concept is sound but impractical because of their enormous complexity.
Preliminary yet partial solutions promise to improve the current situation~\cite{Naehrig:2011:HEP:2046660.2046682}.
To overcome this limitation, several hardware manufacturers extended their architectures with some form of trusted execution environment (TEE).
In a nutshell, a TEE can certify what software it runs, and data stored inside it can only be accessed by its own software.
With TEEs, users do not need to trust the infrastructure provider's execution environment, because it can do no harm to their data, but only the TEE manufacturer.

We use a TEE to ensure the confidentiality and integrity of the \SYS{} proxy.
It is the responsibility of the client to ensure that a certified proxy is running within a trustworthy TEE.
The communication between client and proxy is then encrypted, and the user's real interests are only accessible in the client domain and inside the TEE.
In the following, we present Intel's SGX~\cite{mckeen2013innovative}, our platform of choice and TEE to implement the \SYS{} proxy.

Intel calls an \emph{enclave} a TEE created with SGX.
Enclaves are created and destroyed using specific privileged system calls. 
When an enclave is created, SGX allocates a memory region that is protected from all accesses from outside the enclave itself, including kernel, hypervisor and peripheral DMA.
Applications can interact with enclaves via procedure calls, in both ways.
Parameters and results are copied in and out enclaved memory when a call crosses the enclave border.
Intel offers a software development kit to define and handle in- and out-calls and to manage the enclaves' lifecycle.

The CPU keeps for each enclave a page cache and ensures that each page is assigned to exactly one enclave.
System software, although untrusted, is responsible for assigning pages to enclaves.
An initial set of pages is prepared by the system software, by assigning enclave pages with unencrypted data and code in it.
The CPU keeps a cryptographic hash for the memory pages assigned to each enclave.
After all initial pages are loaded into the enclave, the system software issues an instruction to mark the enclave as initialized.
At that moment the memory hash, or \emph{measurement hash}, is computed.
From this point on, loading unencrypted pages is disabled and application software can enter the protected environment through the enclave interface.

SGX offers instructions for managing keys and for signing certificates of an enclave.
Communication between a remote entity and an enclave is done through a local, untrusted software proxy.
The enclave can send its certificate to the remote entity, which can then verify it with an appropriate authority.
An authentic certificate and a correct measurement hash attest that the correct program has been loaded inside an authentic enclave.
This process is also known as \emph{attestation}.
As certificates are signed within enclaves, remote entities can verify that it was not forged nor modified by an untrusted proxy, and trusted channels can be built (using untrusted components).

Access to enclave memory is prevented by hardware, and all enclaves in a processor can have up to approximately 90MB of a protected memory called EPC (\emph{enclave page cache}).
Paging can still be used to access larger address spaces.
Enclave data residing in the processor's internal cache are hashed and encrypted before flushed to the EPC.
Memory checks are made through a chain of a stateful hash codes using random numbers created every time a page is encrypted.
The chain is stored in untrusted memory, and its root is kept in the CPU, inaccessible from outside, what prevents any tampering attacks in memory, including replay.
Paging is completely handled by untrusted software, in the local operating system.

\subsection{Improving Security with SGX}

SGX has been successfully used to improve the security and privacy of other systems.
Code attestation mechanism coupled with the trusted environment provide an assurance that can enforce security guarantees in a plethora of systems, a few of those described next.

Hoekstra et al.~\cite{hoekstra2013using} show how SGX improves the security of sensitive code and data within three scenarios.
First, they use enclaves in the client-side to store shared secrets with financial institutions, and to generate one-time passwords based on such secrets.
Second, an enterprise-grade digital rights management system that stores document encryption keys within user enclaves. 
Such keys are distributed on demand, and discarded by the enclaves after use.
The documents pass through the enclave for decryption, which in turn generates encrypted bitmaps using the GPU symmetric key.
Third, a video-conferencing application with IP-connected enclaves that exchange encrypted media content and interact with the local hardware using encrypted protocols. These systems prevent malicious software (including high-privilege ones) from gaining access to the private data. 
Verifiable confidential cloud computing (VC3) is a MapReduce implementation with data confidentiality and integrity for both code and data that guarantees that the distributed computation globally ran correctly to completion and was not tampered with~\cite{schuster2015vc3}.
To execute map and reduce tasks, VC3 instantiates enclaves with encrypted code in it.
It implements a key distribution protocol such that guarantees that any enclave that contributes to the job runs the correct code and shares the necessary keys for decrypting code and data.
All data sent to tasks is encrypted, as well as all data produced by the tasks.
Mapper and reducer tasks generate extra encrypted hashes that are used to verify that they properly processed all their input data.
Leveraging enclaves, VC3 supports a threat model with powerful adversaries, that may control all cloud software and hardware, except for the physical processors used in the tasks computations.

\textsc{SCBR} (secure content-based routing) implements a content-based publish/subscribe engine~\cite{pires2016secure} where all message filtering is done inside secure enclaves.
All messages are encrypted when outside enclaves, and the filters operate on plaintext headers.
It uses a hybrid encryption scheme with different keys for header and payload to avoid sending all data through the enclave boundary.
This improves performance and reduces the enclave memory footprint.
An experimental evaluation shows that SCBR adds small overheads when compared to insecure plaintext matching outside enclaves.

Kim at al.~\cite{kim2015first} explored the possibility of using enclaves to provide security and privacy in network applications.
They initially demonstrate how to use enclaves to prevent software-defined inter-domain routers to disclose their routing policies or how the \tor anonymity network~\cite{dingledine2004tor} can be strengthened to run its directory authorities to attest each other.
Attackers can still launch denial-of-service attacks but they cannot alter the directory behavior.
Also, by putting onion routers within enclaves, they can attest their integrity and their admission can be done automatically so directory authorities can be eliminated, and the routers can simply keep track of their membership in a distributed hash table.
Finally, they present how enclaves can be used to securely introduce in-network functionality into TLS sessions.

\textsc{TrustJs} is a framework for trustworthy execution of security-sensitive JavaScript inside commodity browsers~\cite{goltzsche2017trustjs}.
It leverages enclaves to protect the client-side execution of JavaScript, enabling a flexible partitioning of web application code.
Being attested by the server, the enclaved interpreter can be used to offload its computation, which results in lower latencies in the user experience and lower performance demand for the application servers.

Recent work investigate the resilience of SGX enclaves against side-channel attacks~\cite{Weichbrodt2016,xu2015controlled}.
This problem is orthogonal to the one investigated by this paper, and thus considered outside of the scope.

\begin{figure*}[tpb]
\centering
\includegraphics[scale=0.6,trim=0 0 0 0]{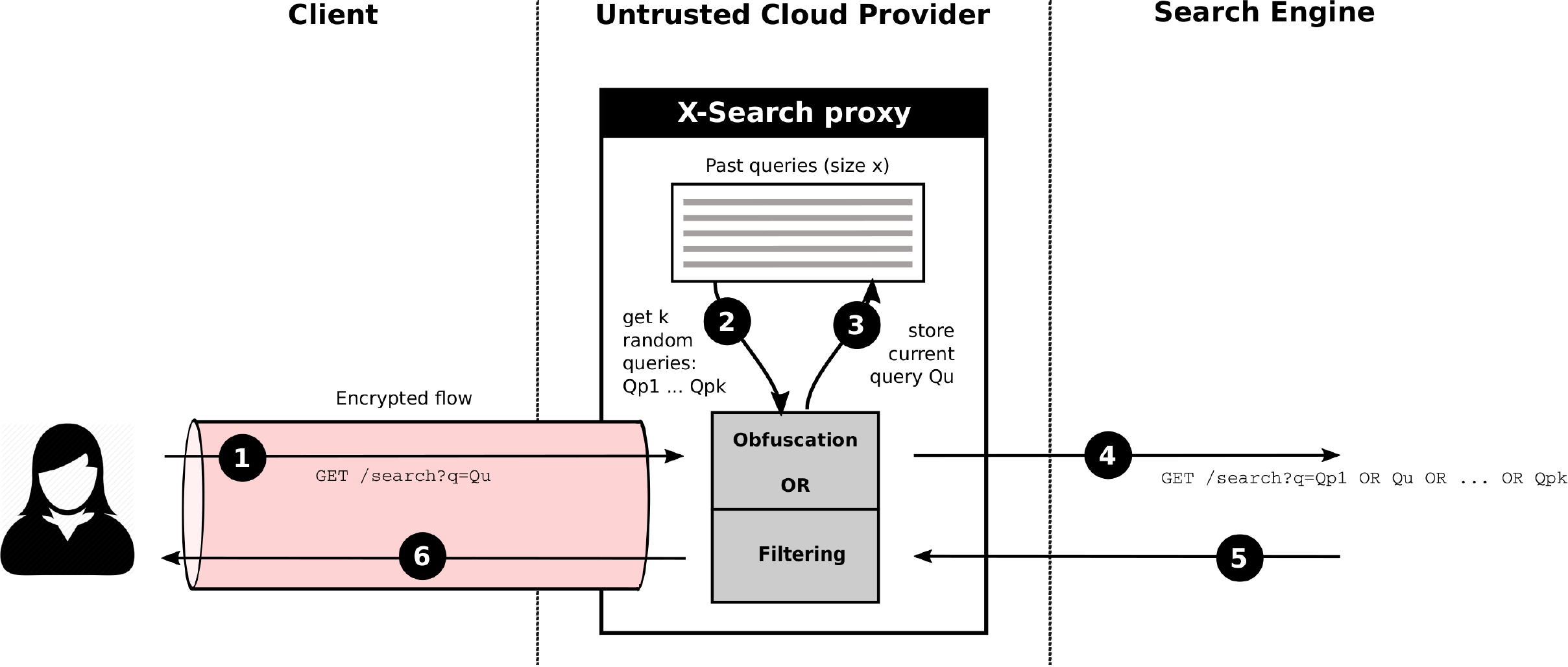}
\caption{The \SYS architecture and execution flow.}
\label{fig:architecture}
\end{figure*}

\section{Adversary model}
\label{sec:adversary}

As further detailed in the following section, the protocol presented in this paper involves three premises:
the client side, the \xsearch proxy nodes running on cloud platforms and the search engine.

We assume that the code and the platform on which client nodes run are trusted.
Then, as further presented in the following section, our protocol relies on \xsearch proxy nodes running on public cloud platforms.
We assume that these nodes are untrusted and can behave in a Byzantine manner~\cite{lamport1982byzantine}, that is they can arbitrarily deviate from a correct behaviour (i.e., they can be subject to a failure, a bug or even behave maliciously). Finally, we assume that the search engine is honest but curious~\cite{Goldreich:2003:CCP:966037.966044}. This means that the search engine behaves correctly when it comes to fetching answers to a specific request but it may collect and exploit in all possible ways the information they receive from clients. 
In particular, we assume that the search engine was able to collect as preliminary information about each user in the system a set of past queries. 
This preliminary information is stored in user profile structures. 

Moreover, we also assume that if the search engine identifies that the client is relying on a private web search mechanism (e.g., an anonymous communication protocol or \SYS), it may run state-of-the-art re-identification attacks (e.g., \cite{Gervais:2014:QWP:2660267.2660367}) in order to re-associate the received request to a known user profile. We further assume that the search engine may collude with proxy nodes (e.g., TOR relays or proxy nodes in \SYS) in order to learn more information about the anonymous client.
 
\section{\xsearch}
\label{sec:xsearch}

We start this section by presenting an overview of our \xsearch protocol (Section~\ref{sec:overview}). Then, we detail how the unlinkability is ensured (Section~\ref{sec:unlink}). Finally, we introduce the obfuscation and filtering mechanisms used to provide indistinguishability (Section~\ref{sec:indis}).

\subsection{Protocol Overview}
\label{sec:overview}

To efficiently protect users during Web search, \xsearch combines unlinkability and indistinguishability.
As previously discussed in Section~\ref{sec:relwork} these two schemes are complementary as the former hides the identity of the requesting user while the latter hides her query.
Figure~\ref{fig:architecture} depicts the architecture and the execution flow of \xsearch.
Specifically, the user interacts with the search engine through an \xsearch proxy node hosted on untrusted public cloud services. We assume the \xsearch proxy to be deployed on physical nodes with available SGX instructions, a scenario that we expect to be common in a near future.

As this proxy node acts as an intermediate node between the search engine and the user, it hides the user identity (i.e., her IP address).
The proxy node is also in charge of obfuscating the user queries, and filtering the results returned from the search engine before forwarding them back to the user.

More precisely, the user starts by sending her query $Q_u$ to the \xsearch proxy (Figure~\ref{fig:architecture} -- \ding{182}).
Then, the proxy node generates a new obfuscated query.
To achieve that, the proxy retrieves $k$ random past queries $Q_{p1},...,Q_{pk}$ (\ding{183}) and aggregates them with the original query in a random order using the logical OR operator.
Next, the proxy stores the initial query in the table of past queries (\ding{184}) and sends one single obfuscated query to the search engine (\ding{185}). 
The search engine is queried by the proxy without using end-to-end encryption~\footnote{Using HTTPS could be also supported by the SGX enclave.}.
Contrary to state of the art indistinguishability protocols, \xsearch uses as fake queries past queries sent by real users. This allows to have fake queries that are effectively indistinguishable from the user's real one. This is possible because past queries are securely stored inside the TEE with no correlation to the identity of their originating users, which prevents any malicious entity from exploiting them.

As the obfuscated query can alter the results returned by the search engine, \emph{e.g.} by mixing results for the original query with results for the additional aggregated past queries, the proxy node includes a filtering step.
Once the search engine sends back the results to the \xsearch proxy (\ding{186}), the filtering removes the results returned by the search engine that are not associated to the original query.
Finally, the remaining results are returned to the user (\ding{187}). These results are tampered by the proxy to remove any URL redirection used for analytics for instance.

We note that the \xsearch proxy node does not maintain individual profile structures associated to each user.
Instead, it only updates a table containing the last $x$ past queries.
To improve performance, the proxy uses multiples threads. The query table is kept in memory and shared among all threads.
Moreover, the user sends her query to the proxy node through an encrypted tunnel with an end point inside the SGX enclave.
Consequently, the protection of the original query is ensured from the client until inside the TEE of the proxy node.
Once outside from the proxy in flight toward the search engine, the original query of the user is protected thanks to the used obfuscation mechanism.

\subsection{Enforcing Unlinkability}
\label{sec:unlink}
The \SYS system offers to end users search unlinkability by relying on a query broker.
This broker runs within the client's domain, such as a local daemon process executing alongside the client's Web browser.
The broker is in charge of the SGX attestation step.
When the user issues a Web search query, her Web client first connects to the local broker.
Then, the broker encrypts the request and forwards the cipher to an \SYS node hosted in an untrusted cloud provider.  
The \SYS node receiving the cipher generates the obfuscated query as further detailed in the following section.
Before sending out the obfuscated query, the original one is securely stored in the SGX reserved memory.
When the search engine sends back the response to the \SYS node, the latter filters out the relevant results, i.e., those related to the original user query, encrypts them and delivers them backward to the broker. 
Finally, the broker decrypts the result and delivers it upward to the Web client.

\subsection{Enforcing Indistinguishability}
\label{sec:indis}

To enforce indistinguishability, \xsearch relies on an obfuscation mechanism.
This mechanism (Algorithm~\ref{algo:generationObuscatedQuery}) aims at hiding the user queries 
among multiple fake queries.
More precisely, the proposed obfuscation mechanism randomly aggregates the original query with $k$
fake queries separated with logical OR operators (lines 2--8).
These fake queries come from the table of past queries maintained in the private memory of the \xsearch proxy (Algorithm~\ref{algo:generationObuscatedQuery}, variable $H$).
Indeed, to avoid building irrelevant fake queries and possibly easily identifiable by the adversary as fake (as discussed in Section~\ref{subsec:issues}), 
the obfuscation mechanism of \xsearch leverages real past queries chosen at random.
Using real past queries ensures that each sub-query of the obfuscated query can be mapped by an adversary conducting a re-identification attack to an existing user profile, thereby making the task of re-identification more complicated to perform.

As an SGX enclave has approximately 90MB of private memory (Section~\ref{subsec:sgx}), we need to bound the memory usage of the \xsearch proxy by limiting the size of $H$ to only keep the $x$ last queries sent by users.
This size limitation acts as a sliding window where only the most recent $x$ queries are exploited.
Once the obfuscated query is generated, the initial query is stored in the history (line 9).

\begin{algorithm}
	\SetKwInOut{Input}{input}
	\Input{$Q$ : initial query, \\
		$H$ : history of queries ($H={Q_0,...,Q_m}$), \\
		$k$ : the number of fake queries. \\}
	$\mathit{obfuscatedQuery} \leftarrow \emptyset$ \;
	$\mathit{index} \leftarrow$ \texttt{\small random}($k+1$) \; \label{algo:fqg:random}
	\While{\texttt{\small sizeof}($\mathit{obfuscatedQuery}$) $<= k$\label{algo:fqg:loop}}{
		\If{$\mathit{index}=0$}{
			$\mathit{obfuscatedQuery} \leftarrow OR(Q)$ \; \label{algo:fqg:pushRealOne}
		}
		\Else{
			$\mathit{obfuscatedQuery} \leftarrow OR(H[$\texttt{\small random}($m$)$])$ \; \label{algo:fqg:pushPastOne}			
		}		
		$\mathit{index} \leftarrow \mathit{index}-1$;
	} 
	$H \leftarrow Q$ \; \label{algo:fqg:updateH}
	\Return $\mathit{obfuscatedQuery}$\;
	\caption{Generation of an obfuscated query}
	\label{algo:generationObuscatedQuery}
\end{algorithm}

This obfuscation mechanism impacts the results returned by the search engine. 
Indeed, the results of the search engine contain a mix of answers corresponding to~$(k+1)$ 
queries (i.e., $k$ fake queries and the initial one).
Consequently, the \xsearch proxy filters the returned results to remove those which are not related to the initial query.
To do this filtering step, the \xsearch node exploits the initial query and the associated fake queries. 
Algorithm~\ref{algo:filtering} describes this filtering process. 
For each result $r$ from the result set, the algorithm determines if it corresponds to the initial query as following. 
A similarity score is assigned to each query (lines 5--6) based on the title and the description of the result. 
The function \texttt{\small nbCommonWords($q, e$)} computes the number of common words between a query $q$ and an element $e$. 
A result $r$ is considered related to the initial query, and hence forwarded to the user, if the initial query has the largest score (lines 7--8).

\begin{algorithm}[thb]
	\SetKwInOut{Input}{input}
	\Input{$Q_u$ : initial query, \\
		$\mathit{pastQuery} = \{Q_{p1}, \dots, Q_{pk}\}$ : set of past queries, \\
		$R$ : set of results for $Q_u \lor Q_{p1} \lor \dots \lor Q_{pk}$. 
	}
	$\bar{R} \leftarrow \emptyset$ \;
	$q^+ \leftarrow \{Q_u, Q_{p1}, \dots, Q_{pk}\}$ \;
	\For{$r \in R$}{
		\For{$q_i \in q^+$}{
			$\mathit{score}[q_i] \leftarrow$ \texttt{\small nbCommonWords}($q_i$, \texttt{\small title}($r$)) \label{algo:nb1} \\
			~ + \texttt{\small nbCommonWords}($q_i$, \texttt{\small desc}($r$))\; \label{algo:nb3}
		}
		\If{$\mathit{score}[Q_u] = \max_{q_i \in q^+} \mathit{score}[q_i]$}{ \label{algo:cond2}
			$\bar{R} \leftarrow \bar{R} \cup \{r\}$ \; \label{algo:ret}
		}
	}
	\Return $\bar{R}$ \;
	\caption{Results filtering.}
	\label{algo:filtering}
\end{algorithm}

 \section{Experimental setup}
\label{sec:experiment}

In this section we present the experimental setup we used to evaluate \xsearch. This comprises: the dataset we used, the comparison baselines we compared against, the evaluation methodology and the metrics used to assess the performance of \xsearch.

\subsection{Web Search Dataset}

To assess \xsearch, we use a real world Web search dataset from the AOL query logs~\cite{pass2006picture}. 
This dataset contains approximately 21 million queries, formulated by 650,000 unique users over three months (from March to May of 2006). 
For the sake of comparison, we use the same methodology as described in~\cite{petit2015peas} to focus our evaluation on the 100 most active users, as they are the most exposed to an adversary willing to unveil their identities.
Indeed, the most active users have exposed more preliminary information to the search engine through their past querying activity.
To reflect this preliminary information collected by the search engine, we built an off-line profile for each user.
To do that, we split the dataset in a training set to build these user profiles, and a testing set to apply and to evaluate the privacy of \xsearch. 
The training set contained two thirds of user queries and the testing set the remaining ones.

\subsection{Comparison Baselines}
We compare the robustness and quality of \SYS against two baselines from the state-of-the art, namely \tor~\cite{dingledine2004tor} and \peas~\cite{petit2015peas}.
As described in Section~\ref{subsubsec:unlink}, \tor leverages a proxy chain to provide unlinkability.
More precisely, this solution uses encryption schemes to hide the identity of a user from the search engine perspective.
\peas, in turn, combines unlinkability and indistinguishability by hiding the identity of the requesting user as well as obfuscating the original query with fake queries.
Specifically, the unlinkability property is ensured by a proxy composed of two trusted nodes relaying the original queries while the obfuscation is achieved locally on the client by aggregating in a random order $k$ fake queries with the original one. These fake queries are generated from the graph of co-occurrence between terms in the history of user queries.
Lastly, we also consider a \direct baseline solution, for which the users send directly their queries to the search engine without any protection.
We do not compare \xsearch against PIR-based solutions because they require to use crypto-based search engines.

\subsection{Methodology}

This section presents the methodology adopted to evaluate \xsearch.
We assess \SYS along three dimensions: the offered privacy (i.e., the protection of users’ queries), the achieved accuracy (i.e., the quality of the results returned by \xsearch), and the pure system performance (i.e., the efficiency of \xsearch in terms of throughput, latency and memory usage).

\subsubsection{Privacy}

To evaluate privacy, we leverage SimAttack~\cite{petit2016simattack} a re-identification attack for which the code is available and that has been shown to outperform previous attacks including a machine learning attack presented in~\cite{peddinti2014web}.
To run this attack, we assume that the attacker holds a set of user profiles built from the learning part of the dataset. Then, we protect each query of the testing part using \xsearch before sending it to the search engine. Then, for each obfuscated query, the attack tries to re-identify both the requesting user and the initial query among fake ones.

More precisely, SimAttack is based on a similarity metric $sim(q, P_u)$ that characterizes the proximity between a query $q$ and a user profile $P_u$.
This profile represents the preliminary information associated to user $u$ collected by the adversary. 
This preliminary information can be viewed as the history of queries of the users before they protect their Web search activities. 
In our case, $P_u$ contains queries that belong to the training set of user $u$.
The similarity metric used by SimAttack accounts the cosine similarity of $q$ and all queries part of the user profile $P_u$, and returns the exponential smoothing of all these similarities ranked in ascending order.
We empirically set the smoothing factor at $0.5$ as it provides the best performances.

To achieve the re-identification from the obfuscated query of \xsearch, we compute the similarity metric for each sub-query embedded in the obfuscated query and each user for which the adversary has a profile.
If only one couple of query and user have the highest similarities, SimAttack returns this couple corresponding to the initial query and to the initial requester.
Otherwise, the attack is unsuccessful.

\subsubsection{Accuracy} 
\sloppy
The obfuscation mechanism of \xsearch (i.e., adding past queries) impacts the results returned by a search engine.
Consequently, we evaluate the capacity of \xsearch to filter results not related to the initial query before forwarding them back to the user.
To achieve that, for a given initial query, we compare results returned by the search engine for this query and the results returned for the associated obfuscated query after the filtering step. 

Our experiments use the Bing search engine.
Search queries are directed to the \texttt{http://www.bing.com/search=q?} address.
As the OR operator implemented by Bing only works with single-word queries, we simulated the execution of an obfuscated query $Q_{obf}\textbf{ }=\textbf{ }Q_{p0}\textbf{ }OR\textbf{ }...\textbf{ }OR\textbf{ }Q_u\textbf{ }OR\textbf{ }...\textbf{ }OR\textbf{ }Q_{pk}$ by submitting each sub-query $Q_{pi}$ and $Q_u$ independently and by merging the $(k + 1)$ result sets.
To circumvent the query$\times$day limit imposed by Bing, for each value of $k$ (i.e., the number of fake queries), we run the experiment on a random subset of the testing set composed of 100 queries. 
Unless otherwise specified, we consider the first 20 results in our accuracy-related experiments.

\subsubsection{Performance}
To evaluate the  performances of \SYS from a system perspective, we implemented a fully-functioning prototype.
Our implementation uses C++ and rely on the Intel SGX SDK (v1.8) libraries and tools~\cite{intel-sdk}.
The prototype is deployed on a machine with an Intel{\textregistered} Core\texttrademark~i7-6700 processor~\cite{intel:i7_6700} and $8\,\mathit{GiB}$ RAM running on \textsc{Ubuntu} 14.04.1 LTS (kernel 4.2.0-42-generic).

The main performance bottlenecks when using intel SGX are known to be the transitions between trusted and untrusted modes (inside/outside the enclaves) and the intensive usage of memory, with two stages: 
(i) when exceeding the processor's last cache level, which requires cache eviction and the consequent cryptographic and integrity checks; 
and (ii) when exceeding the EPC size, triggering memory swaps scheduled by the underlying operating system.
An excessive memory usage can be caused by the management of the past queries inside the enclave's protected memory. 
We evaluate this aspect of \SYS in Section~\ref{sec:evalPerf}.
Furthermore, in order to avoid unnecessary and costly mode transitions, we limit the enclave interface to allow only essential operations that deal with sensitive information. 
Procedure calls made by the vulnerable code are called \emph{ecalls} (enclave calls), whereas the ones made the enclave trusted code are called \emph{ocalls} (outside calls).
The enclave interface offered by the \SYS node is as follows:

\vskip 3mm
\begin{tabular}{p{35mm}p{40mm}}
   \multicolumn{2}{p{0.3\textwidth}}{\textbf{ecalls}} \\[5pt]
   \Call{init}{\emph{ parameters }} &
    Setup options for \SYS{}. \\
   \Call{request}{\emph{ sock, buff, len }} &
    Provision of data to the enclave, coming from the given socket. \\
   \multicolumn{2}{p{0.3\textwidth}}{\textbf{ocalls}} \\[5pt]
   \emph{sock} \Call{connect}{\emph{ host, port }} &
    Performs the DNS lookup and connection to server, returns the socket file descriptor. \\[3pt]
   \Call{send}{\emph{ sock, buff, len }} &
    Sends data through the given socket. \\
   \Call{recv}{\emph{ sock, buff, len }} &
    Receives data from the given socket. \\
   \Call{close}{\emph{ sock }} &
    Close socket file descriptor. \\
\end{tabular}
\vskip 3mm

We measured the system capacity by observing latency for increasing throughput configurations when \SYS{} was configured to reply immediatly to requests.
Memory usage was assessed by populating the past queries store inside the enclave with a real dataset and observing its occupancy.
Finally, we measured respone times considering the complete chain, including the search engine delays.
Results are described in Section~\ref{sec:evalPerf}.

\subsection{Metrics}
\label{sec:experiment:metrics}

We consider three types of metrics in our evaluation. 
The privacy metric measures the level of protection offered by \xsearch and its ability to preserve the users' privacy.
The accuracy metric, in turn, assesses the quality of the query results provided to users according to their original queries.
Lastly, system metrics evaluate the performance and the effectiveness of our solution.

\subsubsection{Privacy}
To assess the privacy we consider the re-identification rate.
This rate aims to retrieve for each protected query, both the content of the initial query 
and the identity of the associated user.
The re-identification rate is defined as follow: 

$$re\textit{-}identification\textbf{ }rate = \frac{|Q_{id}|}{|Q|} $$

where $Q_{id}$ is the set of correctly re-identified queries (i.e., re-identification of both the initial query and the associated user), while $Q$ is the set of original queries sent by users.
This metric is defined between $[0,1]$ where $0$ represents the best solution (i.e., no re-identification) and $1$ represents the worst solution (i.e., all queries are re-identified).

\begin{figure}[t!]
\centering
\includegraphics[scale=0.7,trim=0 0 0 0]{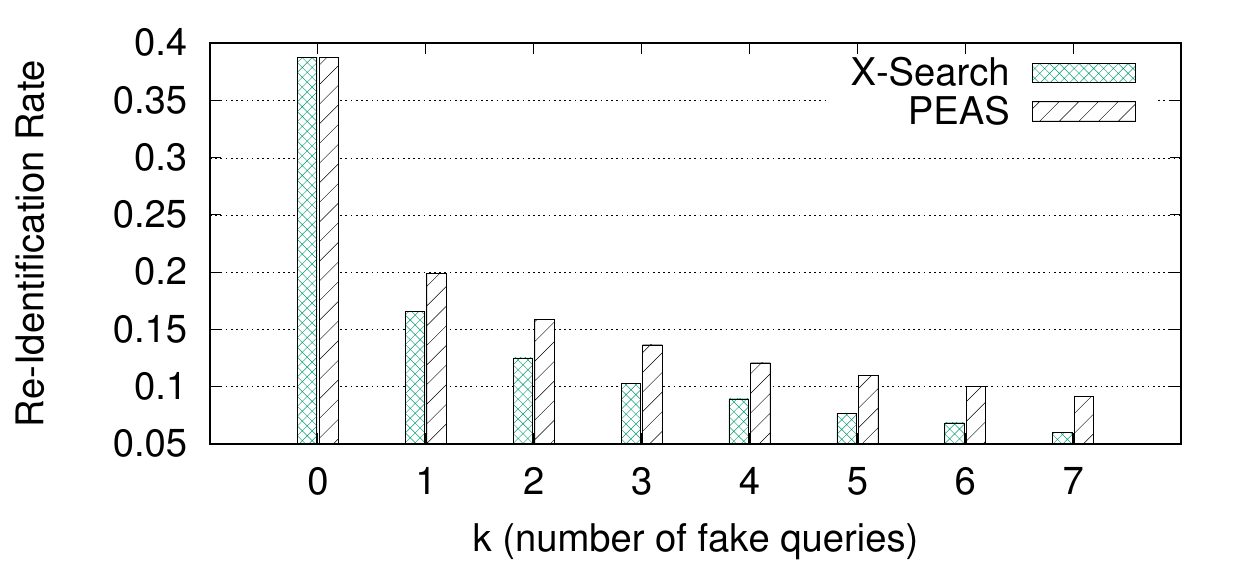}
\caption{X-Search reduces the number of de-anonymized queries compared to PEAS.}
\label{fig:precision-privacy}
\end{figure}

\subsubsection{Accuracy}

The evaluation of the accuracy consists in comparing the lists of results associated to the original query and the results returned with the obfuscated query aggregating the original query and fake ones.
To measure the accuracy, we consider the precision (i.e., correctness) and the recall (i.e., completeness) as defined below:

$$precision = \frac{|R_{or} \cap R_{xs} |}{|R_{xs}|} $$
$$recall = \frac{|R_{or} \cap R_{xs} |}{|R_{or}|} $$

where $R_{or}$ is the set of results returned by the search engine for the original query, and $R_{xs}$ the set of results returned by \xsearch.
Both metrics are in $[0,1]$. The best accuracy is provided with a precision and a recall at $1$.

\subsubsection{System Metrics}

To evaluate the behavior of \xsearch from a systems perspective, we consider the following metrics.
First, we measure the throughput (requests/second) to assess the scalability of \xsearch by measuring its capability to operate properly (adequate response times) even with a growing number of users requesting the service.
Second, looking at occupancy (in MB) using a memory profiler we assess the efficiency of our working prototype. 
Finally, we look at the latency to serve the search results back to the users once they send their queries.

 \section{Evaluation}
\label{sec:evaluation}

This section presents the experimental evaluation of \xsearch 
over three dimensions: the privacy, the accuracy and the system performance, respectively described in Sections~\ref{sec:evalPrivacy},~\ref{sec:evalAccuracy}, and~\ref{sec:evalPerf}.
Our evaluation draws the following conclusions: (1) \xsearch better resists state-of-the-art re-identification attack, (2) it has a limited impact on the accuracy of the results returned to users, and (3) system-wise, it outperforms its competitors, sometimes by orders of magnitude.

\subsection{Privacy}
\label{sec:evalPrivacy}

We start by evaluating the capacity of \xsearch to preserve the user privacy and to improve user protection compared to \peas.
To this end, we measure the robustness of \xsearch against a classical re-identification attack.
Figure~\ref{fig:precision-privacy} shows the re-identification rate for \peas and \xsearch for different values of fake requests, i.e., $k$.
Results for $k=0$ represent the re-identification rate for a solution enforcing only unlinkability (e.g., \tor).
In this case (i.e., without query obfuscation), an adversary using only the history of user queries as preliminary information, is able to re-associate almost $40\%$ of novel queries to their originating user. This confirms that unlikability solutions alone are not sufficient to effectively protect users against re-identification attacks.

Adding only one fake query drops this re-identification rate to $16\%$ for \xsearch and almost $20\%$ for \peas.
This difference comes from the fake query generation process.
Indeed, using real past queries makes \xsearch more robust to the re-identification attack as all sub-queries of the obfuscated query can be mapped to past queries of other users, which creates confusion from the attacker side.
On the contrary, generating fake queries based on the co-occurrence of terms does not ensure \peas to build fake queries closer to a user profile than the original one.

The re-identification rate decreases accordingly to $k$ (i.e., the number of fake queries).
For all value of $k$, \xsearch provides a better protection to the users (i.e., $1 - $\emph{re-identification rate}) than \peas. The improvement of \xsearch over \peas varies from $23\%$ for $k=1$ to $35\%$ for $k=7$.

\begin{figure}[t!]
\centering
\includegraphics[scale=0.7,trim=0 0 0 0]{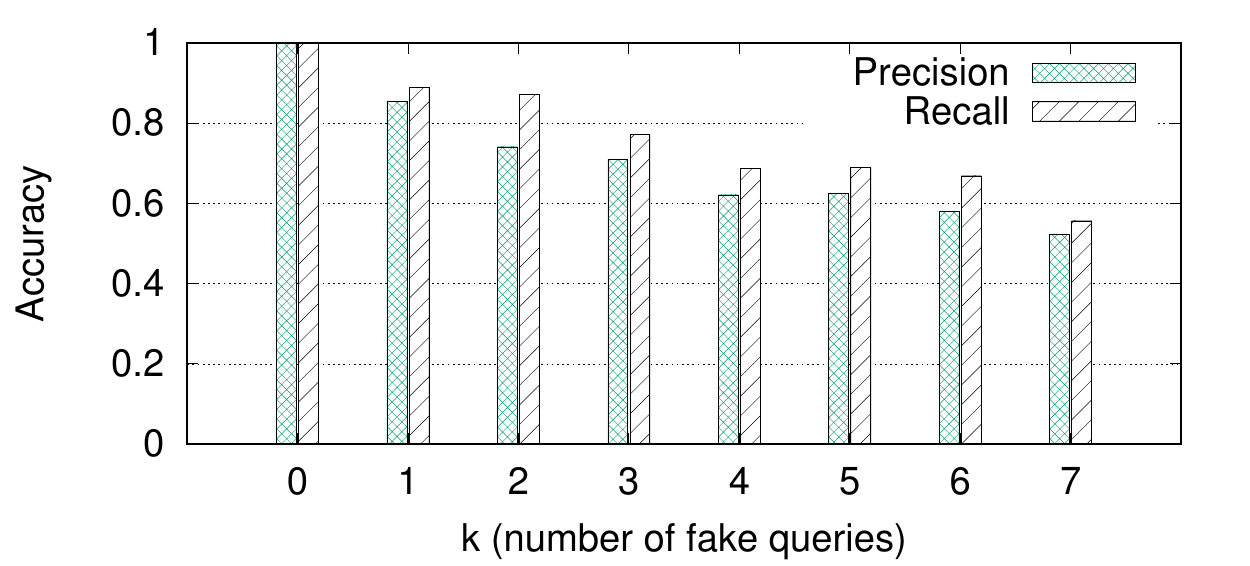}
\caption{Results returned by X-Search are close to results associated to the original query.}
\label{fig:accuracy}
\end{figure}

\subsection{Accuracy}
\label{sec:evalAccuracy}

The accuracy of \xsearch can be measured by evaluating the impact of the obfuscation and the filtering mechanisms on the search results returned to users. 
Specifically, we study if the filtering mechanism is able to remove results related to the fake queries while keeping the ones related to the initial query.
Figure~\ref{fig:accuracy} depicts the precision and the recall of \xsearch according to an increasing value of $k$.
As expected, these curves show that both the recall and the precision slightly decrease according to $k$.
However, the results returned to users are still accurate.
For instance with $k=2$, the value of the recall is higher than $80\%$. 
This means that more than $80\%$ of the results returned to users with \xsearch are the same results as the ones returned if the original query was sent directly to the search engine.
Moreover, the measured precision in this case is higher than $80\%$, which means that only around $20\%$ of the results returned to users can be associated to a fake query and not to the initial query.
These numbers confirm that \xsearch preserves the quality of the results returned by the search engine.

\subsection{System Performance}
\label{sec:evalPerf}
We evaluate the system performance of \SYS to answer the following questions: (1) is our implementation fast? (2) is it memory-efficient and can it be executed within the current SGX memory limitations? and (3) is it usable and responsive to end-users?. 

\begin{figure}[t!]
\centering
\includegraphics[scale=0.7,trim=0 0 0 0]{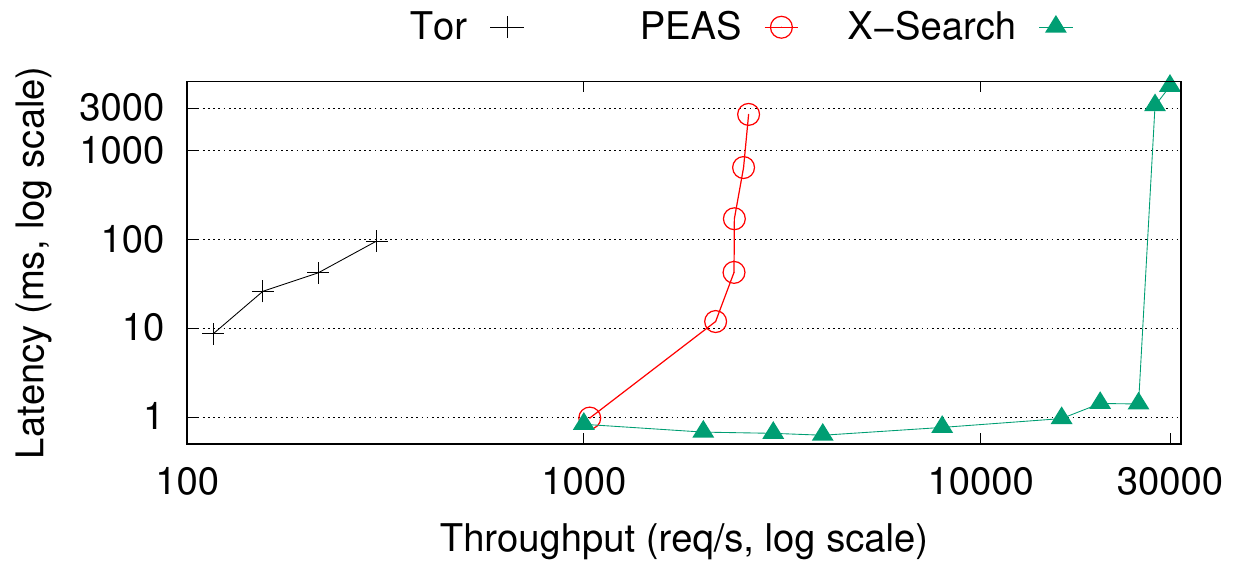}
\caption{Latency/tput rate comparison for \SYS proxy, \peas and \tor.}
\label{fig:xsearch:tputlat}
\end{figure}

We begin by looking at the throughput/latency ratio of the \SYS proxy. 
To perform this experiment, we iteratively increase the rate at which requests are directed toward the \SYS proxy, until the point where the latency to handle each request becomes too high.
For this experiment, we rely on the wrk2 workload generator~\cite{wrk2} to measure the throughput and latency based on the request rates issued to the \SYS proxy.
Note that these measurements are taken without actually hitting the web search engine, to better understand the saturation point of the proxy.
We compare  against \tor and \peas.\footnote{Note that \peas and \tor require custom clients to forge messages following their protocol, whereas \SYS can be used with third-party clients issuing regular HTTP requests, such as \texttt{wget} or \texttt{curl}.}
These results are presented in Figure~\ref{fig:xsearch:tputlat}.
We plot the number of requests per second and the observed latency per request on the x-axis and y-axis, respectively.
Due to the different magnitude of performances, this plot uses a log-log scale.

We observe that \SYS scales well, and it is capable of serving up to $25,000$ requests/sec with sub-second latencies.
Instead, \peas deteriorates much faster, with as few as $1000$ requests/sec being served with a sub-second latency.
In our experiments, \tor performs very poorly: handling as few as 100 requests/sec at an average reply latency of 8.86 milliseconds, around 10$\times$ slower than \SYS serving $1000$ requests/sec.
This result confirms our implementation to be fast and scalable.

\begin{figure}[t!]
\centering
\includegraphics[scale=0.7,trim=0 0 0 0]{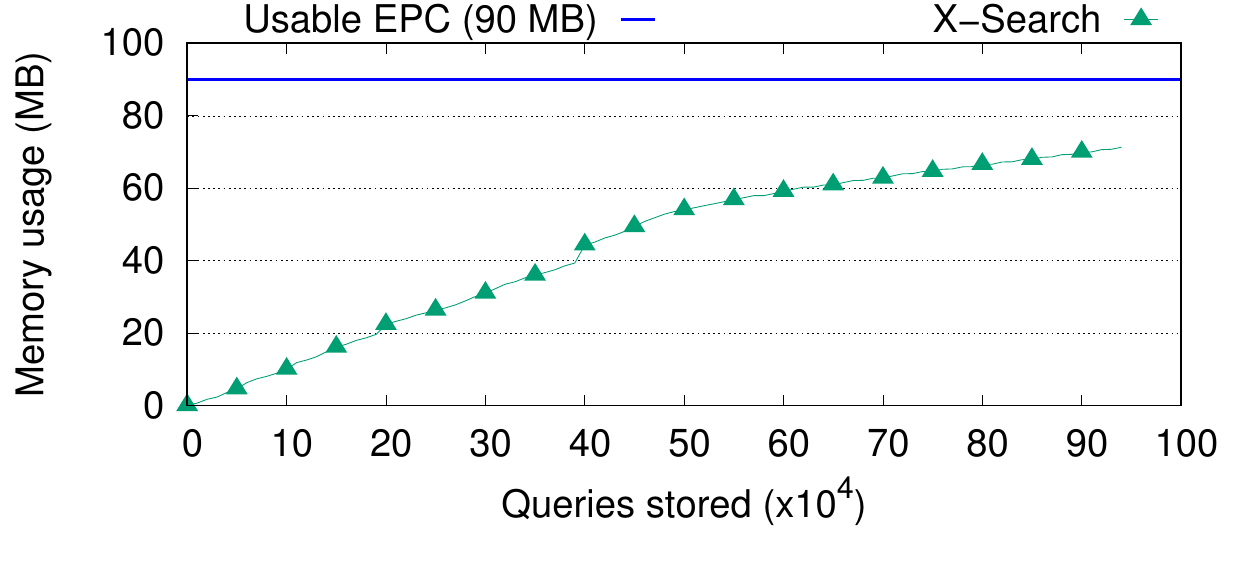}
\caption{X-Search: memory usage. The memory allowed for a single enclave can fit more than 1M queries before hitting the SGX EPC's memory limits.}
\label{fig:xsearch:mem}
\end{figure}

Next, we investigate how much memory is required by the obfuscation scheme.
For this experiment, we used a much larger dataset than the one described in Section~\ref{sec:experiment}. 
Specifically, we use \emph{all} the 6 millions unique queries available in the AOL dataset.
We leverage Valgrind's Massif~\cite{Seward:2008:VAD:1796426} to trace and profile the heap memory allocations executed by the \texttt{xsearch} process.
Figure~\ref{fig:xsearch:mem} presents our result.
Observing the trend of the \SYS curve, it is clear that the EPC size is largely sufficient to store at least 1M queries, a number that can support with ease the obfuscation mechanism.

\begin{figure}[t!]
\centering
\includegraphics[scale=0.7,trim=0 0 0 0]{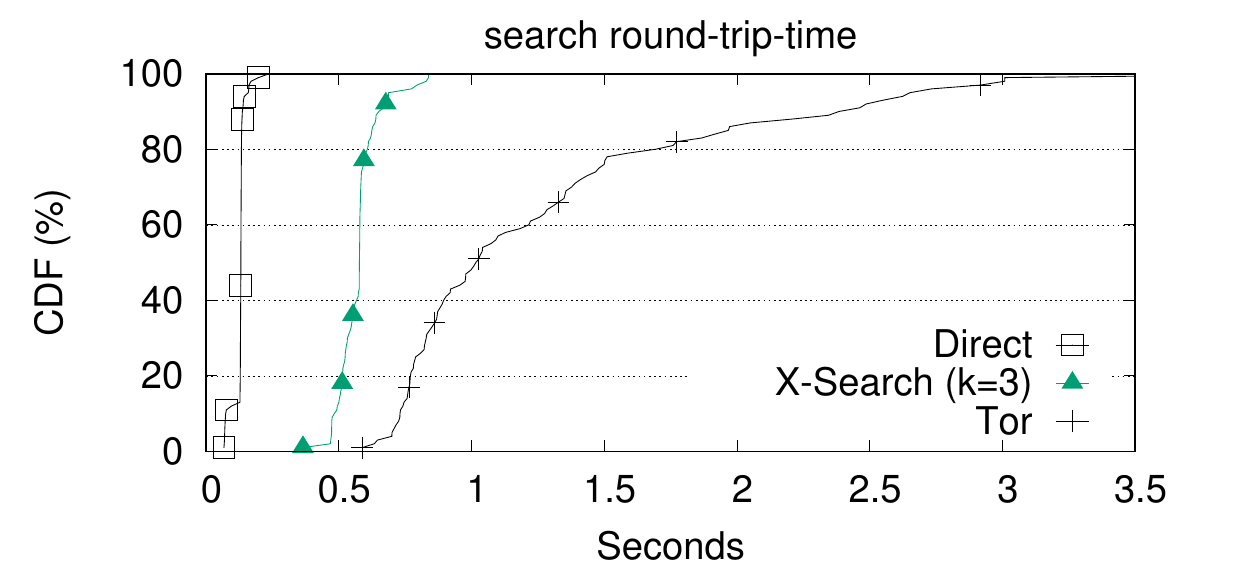}
\caption{User-perceived web search round-trip time for 100 queries with \SYS, over the Tor network and directly contacting the web search engine.}
\label{fig:searchrtt}
\end{figure}

We complete this part of the evaluation by evaluating the user-perceived performance of the system, \emph{e.g.} the end-to-end latency of a Web query from the submission to the reception of the results.
Due to rate limiting schemes adopted by the Bing's search engine, in this experiment we only issue 100 queries,  picked at random between the AOL dataset.
We compare the observed latency between three different scenarios: (1) the client contacting directly the web engine (hence without any privacy guarantees), (2) the same set of queries being routed via the \tor network, and finally (3) using \SYS.
Figure~\ref{fig:searchrtt} presents the results as a Cumulative Distribution Function (CDF) of the measured round-trip network latencies.
We can observe that \SYS allows for much faster replies: the median response time is $0.577$ seconds, and the $99^{th}$ percentile is $0.873$ seconds.
The results over the \tor network are surprisingly bad from a user-perspective: the median time to route a Web search over the onion routers was $1.06$ seconds at the time of our experiments (May 2017), while the $99^{th}$ of the queries complete in up to $3$ seconds.\footnote{We could not conduct a similar experiment using \peas due to a bug in the code.} The \tor network largely exceeds well-known usability margins~\cite{Palmer:2002:WSU:767837.769618}, while \SYS offers a usable and secure browsing experience.

\section{Conclusion}
\label{sec:conclusion}

User behavior tracking by major service providers is one of the main privacy threats in today's Internet.
This is particularly the case with search engines, as they are among the most widely used online services and search queries reveal sensitive information about individual users, such as their age, sex, or religious or political preferences.
Solutions exist in the literature for enabling users to access Web search engines in privacy-preserving way. However, these solutions either do not resist malicious adversaries or are robust but have poor performance. 

In this paper, we proposed a novel architecture for privacy-preserving Web search, which relies on a trusted execution environment (Intel SGX) to support stronger adversarial models than existing solutions.
Our system, \SYS{}, operates as a proxy which stores and leverages user past queries within a protected SGX enclave and generates obfuscated queries on behalf of the user.
It does so by aggregating random past queries in such a way that the search engine is not able to distinguish which one is the original query, but still provides relevant results for the user.
Upon receiving a response from the search engine, the \xsearch proxy filters results to only forward those related to the initial query.

We have implemented a working prototype and evaluated it both analytically and experimentally using real-world datasets.
Our observations indicate that \SYS{} can indeed provide accurate results without disclosing personal information about individual users. Most importantly, \SYS{} does so with a throughput that is orders of magnitude higher than its competitors, i.e., the \peas and \tor protocols.

\section*{Acknowledgments}
The research leading to these results has received funding from the European Commission, Information and Communication Technologies, H2020-ICT-2015 under grant agreement number 690111 (SecureCloud project).
Rafael Pires is also sponsored by CNPq, National Counsel of Technological and Scientific Development, Brazil.

{
\bibliographystyle{ACM-Reference-Format}
\bibliography{biblio}


\begin{thebibliography}{00}


\ifx \showCODEN    \undefined \def \showCODEN     #1{\unskip}     \fi
\ifx \showDOI      \undefined \def \showDOI       #1{{\tt DOI:}\penalty0{#1}\ }
  \fi
\ifx \showISBNx    \undefined \def \showISBNx     #1{\unskip}     \fi
\ifx \showISBNxiii \undefined \def \showISBNxiii  #1{\unskip}     \fi
\ifx \showISSN     \undefined \def \showISSN      #1{\unskip}     \fi
\ifx \showLCCN     \undefined \def \showLCCN      #1{\unskip}     \fi
\ifx \shownote     \undefined \def \shownote      #1{#1}          \fi
\ifx \showarticletitle \undefined \def \showarticletitle #1{#1}   \fi
\ifx \showURL      \undefined \def \showURL       #1{#1}          \fi
\providecommand\bibfield[2]{#2}
\providecommand\bibinfo[2]{#2}
\providecommand\natexlab[1]{#1}
\providecommand\showeprint[2][]{arXiv:#2}

\bibitem[\protect\citeauthoryear{??}{int}{2017}]%
        {intel:i7_6700}
\bibinfo{title}{Intel{\textregistered} {Core{\texttrademark}} {i7-6700}}.
\newblock
  \bibinfo{howpublished}{\url{http://ark.intel.com/products/88196/Intel-Core-i7-6700-Processor-8M-Cache-up-to-4_00-GHz}}.
\newblock


\bibitem[\protect\citeauthoryear{Aguilar-Melchor, Barrier, Fousse, and
  Killijian}{Aguilar-Melchor et~al\mbox{.}}{2016}]%
        {aguilar2016xpir}
\bibfield{author}{\bibinfo{person}{Carlos Aguilar-Melchor},
  \bibinfo{person}{Joris Barrier}, \bibinfo{person}{Laurent Fousse}, {and}
  \bibinfo{person}{Marc-Olivier Killijian}.} \bibinfo{year}{2016}\natexlab{}.
\newblock \showarticletitle{{XPIR: Private information retrieval for
  everyone}}.
\newblock \bibinfo{journal}{{\em Proceedings on Privacy Enhancing
  Technologies\/}}  \bibinfo{volume}{2} (\bibinfo{year}{2016}),
  \bibinfo{pages}{155--174}.
\newblock


\bibitem[\protect\citeauthoryear{Arampatzis, Efraimidis, and
  Drosatos}{Arampatzis et~al\mbox{.}}{2013}]%
        {arampatzis2013query}
\bibfield{author}{\bibinfo{person}{Avi Arampatzis}, \bibinfo{person}{Pavlos~S
  Efraimidis}, {and} \bibinfo{person}{George Drosatos}.}
  \bibinfo{year}{2013}\natexlab{}.
\newblock \showarticletitle{A query scrambler for search privacy on the
  internet}.
\newblock \bibinfo{journal}{{\em Information retrieval\/}}
  \bibinfo{volume}{16}, \bibinfo{number}{6} (\bibinfo{year}{2013}),
  \bibinfo{pages}{657--679}.
\newblock


\bibitem[\protect\citeauthoryear{{Ben Mokhtar}, Berthou, Diarra, Qu{\'e}ma, and
  Shoker}{{Ben Mokhtar} et~al\mbox{.}}{2013}]%
        {mokhtar2013rac}
\bibfield{author}{\bibinfo{person}{Sonia {Ben Mokhtar}},
  \bibinfo{person}{Gautier Berthou}, \bibinfo{person}{Amadou Diarra},
  \bibinfo{person}{Vivien Qu{\'e}ma}, {and} \bibinfo{person}{Ali Shoker}.}
  \bibinfo{year}{2013}\natexlab{}.
\newblock \showarticletitle{Rac: A freerider-resilient, scalable, anonymous
  communication protocol}. In \bibinfo{booktitle}{{\em Distributed Computing
  Systems (ICDCS), 2013 IEEE 33rd International Conference on}}. IEEE,
  \bibinfo{pages}{520--529}.
\newblock


\bibitem[\protect\citeauthoryear{Brickell and Shmatikov}{Brickell and
  Shmatikov}{2006}]%
        {brickell2006efficient}
\bibfield{author}{\bibinfo{person}{Justin Brickell} {and}
  \bibinfo{person}{Vitaly Shmatikov}.} \bibinfo{year}{2006}\natexlab{}.
\newblock \showarticletitle{Efficient anonymity-preserving data collection}. In
  \bibinfo{booktitle}{{\em Proceedings of the 12th ACM SIGKDD international
  conference on Knowledge discovery and data mining}}. ACM,
  \bibinfo{pages}{76--85}.
\newblock


\bibitem[\protect\citeauthoryear{Castelluccia, De~Cristofaro, and
  Perito}{Castelluccia et~al\mbox{.}}{2010}]%
        {castelluccia2010private}
\bibfield{author}{\bibinfo{person}{Claude Castelluccia},
  \bibinfo{person}{Emiliano De~Cristofaro}, {and} \bibinfo{person}{Daniele
  Perito}.} \bibinfo{year}{2010}\natexlab{}.
\newblock \showarticletitle{Private information disclosure from web searches}.
  In \bibinfo{booktitle}{{\em International Symposium on Privacy Enhancing
  Technologies Symposium}}. Springer, \bibinfo{pages}{38--55}.
\newblock


\bibitem[\protect\citeauthoryear{Chaum}{Chaum}{1988}]%
        {chaum1988dining}
\bibfield{author}{\bibinfo{person}{David Chaum}.}
  \bibinfo{year}{1988}\natexlab{}.
\newblock \showarticletitle{The dining cryptographers problem: Unconditional
  sender and recipient untraceability}.
\newblock \bibinfo{journal}{{\em Journal of cryptology\/}} \bibinfo{volume}{1},
  \bibinfo{number}{1} (\bibinfo{year}{1988}), \bibinfo{pages}{65--75}.
\newblock


\bibitem[\protect\citeauthoryear{Corrigan-Gibbs and Ford}{Corrigan-Gibbs and
  Ford}{2010}]%
        {corrigan2010dissent}
\bibfield{author}{\bibinfo{person}{Henry Corrigan-Gibbs} {and}
  \bibinfo{person}{Bryan Ford}.} \bibinfo{year}{2010}\natexlab{}.
\newblock \showarticletitle{Dissent: accountable anonymous group messaging}. In
  \bibinfo{booktitle}{{\em Proceedings of the 17th ACM conference on Computer
  and communications security}}. ACM, \bibinfo{pages}{340--350}.
\newblock


\bibitem[\protect\citeauthoryear{Costan and Devadas}{Costan and Devadas}{}]%
        {costan_intel}
\bibfield{author}{\bibinfo{person}{Victor Costan} {and}
  \bibinfo{person}{Srinivas Devadas}.}
\newblock \bibinfo{booktitle}{{\em {Intel{\textregistered} SGX Explained}}}.
\newblock \bibinfo{type}{{T}echnical {R}eport}.
  \bibinfo{institution}{Cryptology ePrint Archive, Report 2016/086, 2016.}
\newblock


\bibitem[\protect\citeauthoryear{Dingledine, Mathewson, and
  Syverson}{Dingledine et~al\mbox{.}}{2004}]%
        {dingledine2004tor}
\bibfield{author}{\bibinfo{person}{Roger Dingledine}, \bibinfo{person}{Nick
  Mathewson}, {and} \bibinfo{person}{Paul Syverson}.}
  \bibinfo{year}{2004}\natexlab{}.
\newblock \bibinfo{booktitle}{{\em Tor: The second-generation onion router}}.
\newblock \bibinfo{type}{{T}echnical {R}eport}. \bibinfo{institution}{DTIC
  Document}.
\newblock


\bibitem[\protect\citeauthoryear{Domingo-Ferrer, Solanas, and
  Castell{\`a}-Roca}{Domingo-Ferrer et~al\mbox{.}}{2009}]%
        {domingo2009h}
\bibfield{author}{\bibinfo{person}{Josep Domingo-Ferrer},
  \bibinfo{person}{Agusti Solanas}, {and} \bibinfo{person}{Jordi
  Castell{\`a}-Roca}.} \bibinfo{year}{2009}\natexlab{}.
\newblock \showarticletitle{h(k)-Private information retrieval from
  privacy-uncooperative queryable databases}.
\newblock \bibinfo{journal}{{\em Online Information Review\/}}
  \bibinfo{volume}{33}, \bibinfo{number}{4} (\bibinfo{year}{2009}),
  \bibinfo{pages}{720--744}.
\newblock


\bibitem[\protect\citeauthoryear{Gentry}{Gentry}{2009}]%
        {Gentry:2009:FHE:1536414.1536440}
\bibfield{author}{\bibinfo{person}{Craig Gentry}.}
  \bibinfo{year}{2009}\natexlab{}.
\newblock \showarticletitle{Fully Homomorphic Encryption Using Ideal Lattices}.
  In \bibinfo{booktitle}{{\em Proceedings of the Forty-first Annual ACM
  Symposium on Theory of Computing}} {\em (\bibinfo{series}{STOC '09})}.
  \bibinfo{publisher}{ACM}, \bibinfo{address}{New York, NY, USA},
  \bibinfo{pages}{169--178}.
\newblock
\showISBNx{978-1-60558-506-2}
\showDOI{%
\url{https://doi.org/10.1145/1536414.1536440}}


\bibitem[\protect\citeauthoryear{Gervais, Shokri, Singla, Capkun, and
  Lenders}{Gervais et~al\mbox{.}}{2014}]%
        {Gervais:2014:QWP:2660267.2660367}
\bibfield{author}{\bibinfo{person}{Arthur Gervais}, \bibinfo{person}{Reza
  Shokri}, \bibinfo{person}{Adish Singla}, \bibinfo{person}{Srdjan Capkun},
  {and} \bibinfo{person}{Vincent Lenders}.} \bibinfo{year}{2014}\natexlab{}.
\newblock \showarticletitle{Quantifying Web-Search Privacy}. In
  \bibinfo{booktitle}{{\em Proceedings of the 2014 ACM SIGSAC Conference on
  Computer and Communications Security}} {\em (\bibinfo{series}{CCS '14})}.
  \bibinfo{pages}{966--977}.
\newblock


\bibitem[\protect\citeauthoryear{Goldreich}{Goldreich}{2003}]%
        {Goldreich:2003:CCP:966037.966044}
\bibfield{author}{\bibinfo{person}{Oded Goldreich}.}
  \bibinfo{year}{2003}\natexlab{}.
\newblock \showarticletitle{Cryptography and Cryptographic Protocols}.
\newblock \bibinfo{journal}{{\em Distrib. Comput.\/}} \bibinfo{volume}{16},
  \bibinfo{number}{2-3} (\bibinfo{date}{Sept.} \bibinfo{year}{2003}),
  \bibinfo{pages}{177--199}.
\newblock


\bibitem[\protect\citeauthoryear{Goldschlag, Reed, and Syverson}{Goldschlag
  et~al\mbox{.}}{1999}]%
        {goldschlag1999onion}
\bibfield{author}{\bibinfo{person}{David Goldschlag}, \bibinfo{person}{Michael
  Reed}, {and} \bibinfo{person}{Paul Syverson}.}
  \bibinfo{year}{1999}\natexlab{}.
\newblock \showarticletitle{Onion routing}.
\newblock \bibinfo{journal}{{\it Commun. ACM}} \bibinfo{volume}{42},
  \bibinfo{number}{2} (\bibinfo{year}{1999}), \bibinfo{pages}{39--41}.
\newblock


\bibitem[\protect\citeauthoryear{Goltzsche, Wulf, Muthukumaran, Rieck,
  Pietzuch, and Kapitza}{Goltzsche et~al\mbox{.}}{2017}]%
        {goltzsche2017trustjs}
\bibfield{author}{\bibinfo{person}{David Goltzsche}, \bibinfo{person}{Colin
  Wulf}, \bibinfo{person}{Divya Muthukumaran}, \bibinfo{person}{Konrad Rieck},
  \bibinfo{person}{Peter Pietzuch}, {and} \bibinfo{person}{R{\"u}diger
  Kapitza}.} \bibinfo{year}{2017}\natexlab{}.
\newblock \showarticletitle{TrustJS: Trusted Client-side Execution of
  JavaScript}. In \bibinfo{booktitle}{{\em Proceedings of the 10th European
  Workshop on Systems Security}}. ACM, \bibinfo{pages}{7}.
\newblock


\bibitem[\protect\citeauthoryear{Hannak, Sapiezynski, Molavi~Kakhki,
  Krishnamurthy, Lazer, Mislove, and Wilson}{Hannak et~al\mbox{.}}{2013}]%
        {Hannak:2013:MPW:2488388.2488435}
\bibfield{author}{\bibinfo{person}{Aniko Hannak}, \bibinfo{person}{Piotr
  Sapiezynski}, \bibinfo{person}{Arash Molavi~Kakhki},
  \bibinfo{person}{Balachander Krishnamurthy}, \bibinfo{person}{David Lazer},
  \bibinfo{person}{Alan Mislove}, {and} \bibinfo{person}{Christo Wilson}.}
  \bibinfo{year}{2013}\natexlab{}.
\newblock \showarticletitle{{Measuring Personalization of Web Search}}. In
  \bibinfo{booktitle}{{\em Proceedings of the 22Nd International Conference on
  World Wide Web}} {\em (\bibinfo{series}{WWW '13})}. \bibinfo{publisher}{ACM},
  \bibinfo{address}{New York, NY, USA}, \bibinfo{pages}{527--538}.
\newblock
\showISBNx{978-1-4503-2035-1}
\showDOI{%
\url{https://doi.org/10.1145/2488388.2488435}}


\bibitem[\protect\citeauthoryear{Hoekstra, Lal, Pappachan, Phegade, and
  Del~Cuvillo}{Hoekstra et~al\mbox{.}}{2013}]%
        {hoekstra2013using}
\bibfield{author}{\bibinfo{person}{Matthew Hoekstra}, \bibinfo{person}{Reshma
  Lal}, \bibinfo{person}{Pradeep Pappachan}, \bibinfo{person}{Vinay Phegade},
  {and} \bibinfo{person}{Juan Del~Cuvillo}.} \bibinfo{year}{2013}\natexlab{}.
\newblock \showarticletitle{{Using Innovative Instructions to Create
  Trustworthy Software Solutions}}. In \bibinfo{booktitle}{{\em Proceedings of
  the 2Nd International Workshop on Hardware and Architectural Support for
  Security and Privacy}} {\em (\bibinfo{series}{HASP '13})}.
  \bibinfo{publisher}{ACM}, \bibinfo{address}{New York, NY, USA}, Article
  \bibinfo{articleno}{11}, \bibinfo{numpages}{1}~pages.
\newblock
\showISBNx{978-1-4503-2118-1}
\showDOI{%
\url{https://doi.org/10.1145/2487726.2488370}}


\bibitem[\protect\citeauthoryear{Howe and Nissenbaum}{Howe and
  Nissenbaum}{2009}]%
        {howe2009trackmenot}
\bibfield{author}{\bibinfo{person}{Daniel~C Howe} {and} \bibinfo{person}{Helen
  Nissenbaum}.} \bibinfo{year}{2009}\natexlab{}.
\newblock \showarticletitle{TrackMeNot: Resisting surveillance in web search}.
\newblock \bibinfo{journal}{{\em Lessons from the Identity Trail: Anonymity,
  Privacy, and Identity in a Networked Society\/}}  \bibinfo{volume}{23}
  (\bibinfo{year}{2009}), \bibinfo{pages}{417--436}.
\newblock


\bibitem[\protect\citeauthoryear{{Intel Corp.}}{{Intel Corp.}}{2016}]%
        {intel-sdk}
\bibfield{author}{\bibinfo{person}{{Intel Corp.}}}
\newblock
  \bibinfo{howpublished}{\url{https://01.org/intel-software-guard-extensions}}.
\newblock


\bibitem[\protect\citeauthoryear{Kim, Shin, Ha, Kim, and Han}{Kim
  et~al\mbox{.}}{2015}]%
        {kim2015first}
\bibfield{author}{\bibinfo{person}{Seongmin Kim}, \bibinfo{person}{Youjung
  Shin}, \bibinfo{person}{Jaehyung Ha}, \bibinfo{person}{Taesoo Kim}, {and}
  \bibinfo{person}{Dongsu Han}.} \bibinfo{year}{2015}\natexlab{}.
\newblock \showarticletitle{A first step towards leveraging commodity trusted
  execution environments for network applications}. In \bibinfo{booktitle}{{\em
  Proceedings of the 14th ACM Workshop on Hot Topics in Networks}}. ACM,
  \bibinfo{pages}{7}.
\newblock


\bibitem[\protect\citeauthoryear{Lamport, Shostak, and Pease}{Lamport
  et~al\mbox{.}}{1982}]%
        {lamport1982byzantine}
\bibfield{author}{\bibinfo{person}{Leslie Lamport}, \bibinfo{person}{Robert
  Shostak}, {and} \bibinfo{person}{Marshall Pease}.}
  \bibinfo{year}{1982}\natexlab{}.
\newblock \showarticletitle{The Byzantine generals problem}.
\newblock \bibinfo{journal}{{\em ACM Transactions on Programming Languages and
  Systems (TOPLAS)\/}} \bibinfo{volume}{4}, \bibinfo{number}{3}
  (\bibinfo{year}{1982}), \bibinfo{pages}{382--401}.
\newblock


\bibitem[\protect\citeauthoryear{Langville and Meyer}{Langville and
  Meyer}{2011}]%
        {langville2011google}
\bibfield{author}{\bibinfo{person}{Amy~N Langville} {and}
  \bibinfo{person}{Carl~D Meyer}.} \bibinfo{year}{2011}\natexlab{}.
\newblock \bibinfo{booktitle}{{\em Google's PageRank and beyond: The science of
  search engine rankings}}.
\newblock \bibinfo{publisher}{Princeton University Press}.
\newblock


\bibitem[\protect\citeauthoryear{Lindell and Waisbard}{Lindell and
  Waisbard}{2010}]%
        {lindell2010private}
\bibfield{author}{\bibinfo{person}{Yehuda Lindell} {and} \bibinfo{person}{Erez
  Waisbard}.} \bibinfo{year}{2010}\natexlab{}.
\newblock \showarticletitle{Private web search with malicious adversaries}. In
  \bibinfo{booktitle}{{\em International Symposium on Privacy Enhancing
  Technologies Symposium}}. Springer, \bibinfo{pages}{220--235}.
\newblock


\bibitem[\protect\citeauthoryear{McKeen, Alexandrovich, Berenzon, Rozas, Shafi,
  Shanbhogue, and Savagaonkar}{McKeen et~al\mbox{.}}{2013}]%
        {mckeen2013innovative}
\bibfield{author}{\bibinfo{person}{Frank McKeen}, \bibinfo{person}{Ilya
  Alexandrovich}, \bibinfo{person}{Alex Berenzon}, \bibinfo{person}{Carlos~V.
  Rozas}, \bibinfo{person}{Hisham Shafi}, \bibinfo{person}{Vedvyas Shanbhogue},
  {and} \bibinfo{person}{Uday~R. Savagaonkar}.}
  \bibinfo{year}{2013}\natexlab{}.
\newblock \showarticletitle{Innovative instructions and software model for
  isolated execution}. In \bibinfo{booktitle}{{\em {HASP} 2013, The Second
  Workshop on Hardware and Architectural Support for Security and Privacy,
  Tel-Aviv, Israel, June 23-24, 2013}}. \bibinfo{pages}{10}.
\newblock
\showDOI{%
\url{https://doi.org/10.1145/2487726.2488368}}


\bibitem[\protect\citeauthoryear{Naehrig, Lauter, and Vaikuntanathan}{Naehrig
  et~al\mbox{.}}{2011}]%
        {Naehrig:2011:HEP:2046660.2046682}
\bibfield{author}{\bibinfo{person}{Michael Naehrig}, \bibinfo{person}{Kristin
  Lauter}, {and} \bibinfo{person}{Vinod Vaikuntanathan}.}
  \bibinfo{year}{2011}\natexlab{}.
\newblock \showarticletitle{Can Homomorphic Encryption Be Practical?}. In
  \bibinfo{booktitle}{{\em Proceedings of the 3rd ACM Workshop on Cloud
  Computing Security Workshop}} {\em (\bibinfo{series}{CCSW '11})}.
  \bibinfo{publisher}{ACM}, \bibinfo{address}{New York, NY, USA},
  \bibinfo{pages}{113--124}.
\newblock
\showISBNx{978-1-4503-1004-8}
\showDOI{%
\url{https://doi.org/10.1145/2046660.2046682}}


\bibitem[\protect\citeauthoryear{Palmer}{Palmer}{2002}]%
        {Palmer:2002:WSU:767837.769618}
\bibfield{author}{\bibinfo{person}{Jonathan~W. Palmer}.}
  \bibinfo{year}{2002}\natexlab{}.
\newblock \showarticletitle{Web Site Usability, Design, and Performance
  Metrics}.
\newblock \bibinfo{journal}{{\em Info. Sys. Research\/}} \bibinfo{volume}{13},
  \bibinfo{number}{2} (\bibinfo{date}{June} \bibinfo{year}{2002}),
  \bibinfo{pages}{151--167}.
\newblock
\showISSN{1526-5536}
\showDOI{%
\url{https://doi.org/10.1287/isre.13.2.151.88}}


\bibitem[\protect\citeauthoryear{Pang, Shen, and Krishnan}{Pang
  et~al\mbox{.}}{2010}]%
        {pang2010privacy}
\bibfield{author}{\bibinfo{person}{Hweehwa Pang}, \bibinfo{person}{Jialie
  Shen}, {and} \bibinfo{person}{Ramayya Krishnan}.}
  \bibinfo{year}{2010}\natexlab{}.
\newblock \showarticletitle{Privacy-preserving similarity-based text
  retrieval}.
\newblock \bibinfo{journal}{{\em ACM Transactions on Internet Technology
  (TOIT)\/}} \bibinfo{volume}{10}, \bibinfo{number}{1} (\bibinfo{year}{2010}),
  \bibinfo{pages}{4}.
\newblock


\bibitem[\protect\citeauthoryear{Pass, Chowdhury, and Torgeson}{Pass
  et~al\mbox{.}}{2006}]%
        {pass2006picture}
\bibfield{author}{\bibinfo{person}{Greg Pass}, \bibinfo{person}{Abdur
  Chowdhury}, {and} \bibinfo{person}{Cayley Torgeson}.}
  \bibinfo{year}{2006}\natexlab{}.
\newblock \showarticletitle{{A Picture of Search}}. In \bibinfo{booktitle}{{\em
  Proceedings of the 1st International Conference on Scalable Information
  Systems}} {\em (\bibinfo{series}{InfoScale '06})}. \bibinfo{publisher}{ACM},
  \bibinfo{address}{New York, NY, USA}, Article \bibinfo{articleno}{1}.
\newblock
\showISBNx{1-59593-428-6}
\showDOI{%
\url{https://doi.org/10.1145/1146847.1146848}}


\bibitem[\protect\citeauthoryear{Peddinti and Saxena}{Peddinti and
  Saxena}{2014}]%
        {peddinti2014web}
\bibfield{author}{\bibinfo{person}{Sai~Teja Peddinti} {and}
  \bibinfo{person}{Nitesh Saxena}.} \bibinfo{year}{2014}\natexlab{}.
\newblock \showarticletitle{Web search query privacy: Evaluating query
  obfuscation and anonymizing networks1}.
\newblock \bibinfo{journal}{{\em Journal of Computer Security\/}}
  \bibinfo{volume}{22}, \bibinfo{number}{1} (\bibinfo{year}{2014}),
  \bibinfo{pages}{155--199}.
\newblock


\bibitem[\protect\citeauthoryear{Petit, Cerqueus, Boutet, Mokhtar, Coquil,
  Brunie, and Kosch}{Petit et~al\mbox{.}}{2016}]%
        {petit2016simattack}
\bibfield{author}{\bibinfo{person}{Albin Petit}, \bibinfo{person}{Thomas
  Cerqueus}, \bibinfo{person}{Antoine Boutet}, \bibinfo{person}{Sonia~Ben
  Mokhtar}, \bibinfo{person}{David Coquil}, \bibinfo{person}{Lionel Brunie},
  {and} \bibinfo{person}{Harald Kosch}.} \bibinfo{year}{2016}\natexlab{}.
\newblock \showarticletitle{{SimAttack}: Private Web Search under Fire}.
\newblock \bibinfo{journal}{{\em Journal of Internet Services and
  Applications\/}} \bibinfo{volume}{7}, \bibinfo{number}{1}
  (\bibinfo{year}{2016}), \bibinfo{pages}{2}.
\newblock


\bibitem[\protect\citeauthoryear{Petit, Cerqueus, Mokhtar, Brunie, and
  Kosch}{Petit et~al\mbox{.}}{2015}]%
        {petit2015peas}
\bibfield{author}{\bibinfo{person}{Albin Petit}, \bibinfo{person}{Thomas
  Cerqueus}, \bibinfo{person}{Sonia~Ben Mokhtar}, \bibinfo{person}{Lionel
  Brunie}, {and} \bibinfo{person}{Harald Kosch}.}
  \bibinfo{year}{2015}\natexlab{}.
\newblock \showarticletitle{{PEAS}: Private, Efficient and Accurate Web
  Search}. In \bibinfo{booktitle}{{\em Trustcom/BigDataSE/ISPA, 2015 IEEE}},
  Vol.~\bibinfo{volume}{1}. IEEE, \bibinfo{pages}{571--580}.
\newblock


\bibitem[\protect\citeauthoryear{Pires, Pasin, Felber, and Fetzer}{Pires
  et~al\mbox{.}}{2016}]%
        {pires2016secure}
\bibfield{author}{\bibinfo{person}{Rafael Pires}, \bibinfo{person}{Marcelo
  Pasin}, \bibinfo{person}{Pascal Felber}, {and} \bibinfo{person}{Christof
  Fetzer}.} \bibinfo{year}{2016}\natexlab{}.
\newblock \showarticletitle{Secure Content-Based Routing Using Intel Software
  Guard Extensions}. In \bibinfo{booktitle}{{\em Proceedings of the 17th
  International Middleware Conference}} {\em (\bibinfo{series}{Middleware
  '16})}. ACM, \bibinfo{pages}{10}.
\newblock
\showDOI{%
\url{https://doi.org/10.1145/2988336.2988346}}


\bibitem[\protect\citeauthoryear{Schuster, Costa, Fournet, Gkantsidis, Peinado,
  Mainar-Ruiz, and Russinovich}{Schuster et~al\mbox{.}}{2015}]%
        {schuster2015vc3}
\bibfield{author}{\bibinfo{person}{Felix Schuster}, \bibinfo{person}{Manuel
  Costa}, \bibinfo{person}{C{\'e}dric Fournet}, \bibinfo{person}{Christos
  Gkantsidis}, \bibinfo{person}{Marcus Peinado}, \bibinfo{person}{Gloria
  Mainar-Ruiz}, {and} \bibinfo{person}{Mark Russinovich}.}
  \bibinfo{year}{2015}\natexlab{}.
\newblock \showarticletitle{VC3: Trustworthy data analytics in the cloud using
  SGX}. In \bibinfo{booktitle}{{\em Security and Privacy (SP), 2015 IEEE
  Symposium on}}. IEEE, \bibinfo{pages}{38--54}.
\newblock


\bibitem[\protect\citeauthoryear{Seward, Nethercote, and Weidendorfer}{Seward
  et~al\mbox{.}}{2008}]%
        {Seward:2008:VAD:1796426}
\bibfield{author}{\bibinfo{person}{J. Seward}, \bibinfo{person}{N. Nethercote},
  {and} \bibinfo{person}{J. Weidendorfer}.} \bibinfo{year}{2008}\natexlab{}.
\newblock \bibinfo{booktitle}{{\em Valgrind 3.3 - Advanced Debugging and
  Profiling for GNU/Linux Applications}}.
\newblock \bibinfo{publisher}{Network Theory Ltd.}
\newblock
\showISBNx{0954612051, 9780954612054}


\bibitem[\protect\citeauthoryear{Weichbrodt, Kurmus, Pietzuch, and
  Kapitza}{Weichbrodt et~al\mbox{.}}{2016}]%
        {Weichbrodt2016}
\bibfield{author}{\bibinfo{person}{Nico Weichbrodt}, \bibinfo{person}{Anil
  Kurmus}, \bibinfo{person}{Peter Pietzuch}, {and} \bibinfo{person}{R{\"u}diger
  Kapitza}.} \bibinfo{year}{2016}\natexlab{}.
\newblock \bibinfo{booktitle}{{\em AsyncShock: Exploiting Synchronisation Bugs
  in Intel SGX Enclaves}}.
\newblock \bibinfo{publisher}{Springer International Publishing},
  \bibinfo{address}{Cham}, \bibinfo{pages}{440--457}.
\newblock
\showISBNx{978-3-319-45744-4}
\showDOI{%
\url{https://doi.org/10.1007/978-3-319-45744-4_22}}


\bibitem[\protect\citeauthoryear{Wolinsky, Corrigan-Gibbs, Ford, and
  Johnson}{Wolinsky et~al\mbox{.}}{2012}]%
        {wolinsky2012dissent}
\bibfield{author}{\bibinfo{person}{David~Isaac Wolinsky},
  \bibinfo{person}{Henry Corrigan-Gibbs}, \bibinfo{person}{Bryan Ford}, {and}
  \bibinfo{person}{Aaron Johnson}.} \bibinfo{year}{2012}\natexlab{}.
\newblock \showarticletitle{Dissent in Numbers: Making Strong Anonymity
  Scale.}. In \bibinfo{booktitle}{{\em OSDI}}. \bibinfo{pages}{179--182}.
\newblock


\bibitem[\protect\citeauthoryear{wrk2}{wrk2}{2015}]%
        {wrk2}
wrk2.
\newblock \bibinfo{title}{{Wrk2.}}
\newblock \bibinfo{howpublished}{\url{https://github.com/giltene/wrk2}}.
\newblock


\bibitem[\protect\citeauthoryear{Xu, Cui, and Peinado}{Xu
  et~al\mbox{.}}{2015}]%
        {xu2015controlled}
\bibfield{author}{\bibinfo{person}{Yuanzhong Xu}, \bibinfo{person}{Weidong
  Cui}, {and} \bibinfo{person}{Marcus Peinado}.}
  \bibinfo{year}{2015}\natexlab{}.
\newblock \showarticletitle{Controlled-channel attacks: Deterministic side
  channels for untrusted operating systems}. In \bibinfo{booktitle}{{\em
  Security and Privacy (SP), 2015 IEEE Symposium on}}. IEEE,
  \bibinfo{pages}{640--656}.
\newblock


\bibitem[\protect\citeauthoryear{Yang and Ghose}{Yang and Ghose}{2010}]%
        {yang2010analyzing}
\bibfield{author}{\bibinfo{person}{Sha Yang} {and} \bibinfo{person}{Anindya
  Ghose}.} \bibinfo{year}{2010}\natexlab{}.
\newblock \showarticletitle{Analyzing the relationship between organic and
  sponsored search advertising: Positive, negative, or zero interdependence?}
\newblock \bibinfo{journal}{{\em Marketing Science\/}} \bibinfo{volume}{29},
  \bibinfo{number}{4} (\bibinfo{year}{2010}), \bibinfo{pages}{602--623}.
\newblock


\end{thebibliography}
}

\end{document}